\documentclass{aa}
\usepackage{graphics}
\usepackage{epsfig}
\begin{document}

   \thesaurus{11     
              (11.09.1;  
               11.12.1;  
               11.06.2;  
               11.19.3;  
               11.04.2;} 
%
   \title{IC 10: More evidence that it is a blue compact dwarf}

   \subtitle{}

   \author{
          M. G. Richer \inst{1}
          \and A. Bullejos \inst{2,3}
          \and J. Borissova \inst{4}
          \and Marshall L. McCall \inst{5}
          \and Henry Lee \inst{5}
          \and R. Kurtev \inst{6}
          \and L. Georgiev \inst{3}
          \and R. L. Kingsburgh \inst{5}
          \and R. Ross \inst{5}
          \and M. Rosado \inst{3}
          }

   \offprints{M. G. Richer}

   \institute{Observatorio Astron\'omico Nacional,
              Instituto de Astronom\'\i a, UNAM, P. O. Box 439027, San Diego, CA
              92143-9027 \\ (richer@astrosen.unam.mx)
              \and
              Instituto de Astrof\'\i sica de Canarias, V\'\i a
              L\'actea, E-38200 La Laguna, Tenerife, Spain
              (almudena@astroscu.unam.mx)
              \and
              Instituto de Astronom\'\i a, UNAM, Apartado Postal 70-264,
              Ciudad Universitaria, 04510 M\'exico, D. F.,
              M\'exico \\
              (georgiev@astroscu.unam.mx,
              margarit@astroscu.unam.mx)
              \and
              Institute of Astronomy, Bulgarian Academy of
              Sciences and Isaac Newton Institute of Chile, Bulgarian Branch,
              72 Tsarigradsko chauss\`ee, BG-1784 Sofia,
              Bulgaria  (jura@haemimont.bg)
              \and
              Dept. of Physics \& Astronomy, York University, 4700
              Keele Street, Toronto, Ontario, Canada   M3J 1P3 \\
              (mccall@aries.phys.yorku.ca,
              lee@aries.phys.yorku.ca,
              robin@aries.phys.yorku.ca)
              \and
              Department of Astronomy and Isaac Newton Institute of Chile,
              Bulgarian Branch, Sofia University, BG-1164
              Sofia, Bulgaria (kurtev@phys.uni-sofia.bg)
              }

   \date{Received ; }

   \maketitle

   \begin{abstract}

   We present optical spectroscopy of \ion{H}{ii} regions in the
   Local Group galaxy IC 10 and $UBVR$ photometry of foreground
   stars in three fields towards this galaxy.  From these data, we
   find that the foreground reddening due to the Milky Way is
   $E(B-V)=0.77\pm 0.07\,\mathrm{mag}$.
   We find that IC 10 contains considerable internal dust, which qualitatively
   explains the variety of reddening values found by studies of its
   different stellar populations.
   Based upon our foreground reddening, IC 10 has
   intrinsic photometric properties like those of a blue compact dwarf
   galaxy, and not those of a dwarf irregular.  This result is
   consistent with much evidence that IC 10 is in the throes of a
   starburst that began at least $10\,\mathrm{Myr}$ ago.
   We also
   report the discovery of a new WR star in the \ion{H}{ii} region
   HL111c.

   \end{abstract}

%

\section{Introduction}

IC 10 has long been recognized as a peculiar object (Hubble
\cite{Hubble1936}), and is now considered the Local Group's
representative of a starburst galaxy (van den Bergh
\cite{vandenBergh2000}). As we will show, IC 10 may very well be
the nearest example of a blue compact dwarf galaxy (BCD).
Certainly, its surface brightness is similar to that seen in BCDs
once we account for the foreground reddening. Among Local Group
dwarf galaxies, IC 10 has the highest surface density of WR stars
and the highest current rate of star formation (Mateo
\cite{Mateo1998}). The presence of so many WR stars and the high
H$\alpha$ luminosity emphasize that IC 10 is undergoing a strong
burst of star formation that began at least 10\,Myr ago.
Observations of 21-cm emission from \ion{H}{i} reveal that IC 10
consists of an inner disk embedded in an extended, complex,
counter-rotating envelope (Shostak \& Skillman
\cite{ShostakSkillman1989}), and leads Wilcots \& Miller
(\cite{WilcotsMiller1998}) to conclude that IC 10 is still in its
formative stage. These \ion{H}{i} observations emphasize the youth
of the current star formation episode, for there is a notable lack
of interstellar medium structures that are attributable to
supernovae (Wilcots \& Miller \cite{WilcotsMiller1998}).  Several
studies of the stellar populations in IC 10 have revealed the
existence of young, intermediate-age, and old stellar populations
(Massey \& Armandroff \cite{MasseyArmandroff1995}; Sakai et al.
\cite{Sakaietal1999}; Borissova et al. \cite{Borissovaetal2000}).
However, with the exception of the very recent star formation,
very little is known of the history of star formation in IC 10,
and {\it nothing} is known of the star formation history outside
of the star-forming region.

All optical studies of IC 10 are hampered by the large foreground
reddening due to its position near the plane of our galaxy $(l,b)
= (119^\circ, -3^\circ)$.  Our objective here is to reconsider the
nature of IC 10 on the bases of its photometric properties and a
new determination of its foreground reddening.  We investigate its
foreground reddening using spectroscopy of eleven of its
\ion{H}{ii} regions. We find that IC 10 should be considered a BCD
instead of a typical dwarf irregular galaxy.

\section{Observations and data reduction}

\begin{table*}
\caption[]{Observations log} \label{Obslog}
\[
\begin{tabular}{llllll}
\hline \noalign{\smallskip}
Observatory & Steward long slit & SPM long slit
& GH long slit & GH multi-object & Rozhen \\
Telescope & 2.3 m & 2.1m & 2.1m & 2.1m & 2m  \\
\noalign{\smallskip} \hline \noalign{\smallskip}
Date & 14 Oct 1991 & 1 Dec 1994 & 7 Jan 1999 & 8-9 Jan 1999 & 14-16 Sep 1999   \\
CCD & Texas Instruments & Tektronix & Tektronix & EEV & Tektronix \\
CCD format & $800 \times 800$ & $1024 \times 1024$ & $1024 \times 1024$ & $576 \times
384$ & $1024 \times 1024$  \\
gain ($e^{-}$/ADU) & 2.8 & 1.22 & 1.85 & & 4.93 \\
read noise ($e^{-}$) & 7.8 & 3.0 & 3.7 & 8.0 & 5.1 \\
disperser & grating & grating & grating & grism &  \\
ruling (l/mm) & $600$  & $300$ & $300$ & &  \\
$\lambda_\mathrm{blaze}$ (\AA)  & $3568$ & $5000$ & $6693$ & $5500$ & \\
order & first & first & first & first & \\
$\lambda$ range (\AA)& $3621-5134$& $3450-7450$ & $3680-6966$ & $3700-6900$ & $UBVR$   \\
dispersion (\AA/pix) & $1.89$ & $4.00$ & $3.21$ & $5.61$ \\
arc lamp& He-Ar & He-Ar & He-Ar & Ne-Ar \\
standard stars & Hiltner102 & Hiltner102 & G191-B2B  & HD19445 & Landolt (1992) \\
 & LTT9239 & G191-B2B &  HD19445 & HD74721 & standard fields   \\
 & BD+284211 & HD217086 & HD109995 & &    \\
 & Hiltner600 & HD849973 & & &  \\

 \noalign{\smallskip} \hline \noalign{\smallskip}
\end{tabular}
\]
\end{table*}

\subsection{Spectroscopy}

The spectroscopic observations were carried out at three different
observatories. The details of the observations are listed in Table
\ref{Obslog}. At Steward Observatory, the spectra were obtained
with the Boller \& Chivens (B\&C) grating spectrograph at the
telescope's Cassegrain focus.  The spectrograph slit covered
$2.\!\arcsec 5  \times 4.\!\arcmin 1$ on the sky and was oriented
at a position angle of 10\degr\ (from north towards east). At the
Observatorio Astron\'omico Nacional in San Pedro M\'artir (SPM),
Baja California, M\'exico, the spectra were obtained with a
similar B\&C spectrograph at the telescope's Cassegrain focus. The
spectrograph slit covered $2.\!\arcsec 2 \times 5\arcmin$ on the
sky and was oriented and positioned very similarly to the Steward
observations. At the Observatorio Astrof\'\i sico Guillermo Haro
(GH), Sonora, M\'exico, the observations were obtained using two
techniques. First, we obtained long slit spectroscopy using a B\&C
spectrograph at the telescope's Cassegrain focus. The slit covered
$2.\!\arcsec 5 \times 3\arcmin$ on the sky and was oriented
east-west. Second, we obtained multi-object spectroscopy with the
LFOSC imaging spectrograph, which uses a transmission grism as the
dispersing element (Zickgraf et al. \cite{Zickgrafetal1997}).
Objects were selected for spectroscopy using focal plane masks
constructed from direct images obtained with the spectrograph.
Holes of 3\arcsec\ in size were used for the objects of interest.
Any object could be selected provided that it was not aligned with
another in declination, for the dispersion axis was oriented
east-west. The standard stars that were observed in each case are
listed in Table \ref{Obslog}.

The reduction of these spectra was carried out using the Image
Reduction and Analysis Facility (IRAF)\footnote{IRAF is
distributed by the National Optical Astronomical Observatories,
which is operated by the Associated Universities for Research in
Astronomy, Inc., under contract to the National Science
Foundation.} software package (specifically the specred package).
First, the bias images were combined and their signature
subtracted from all of the images. Next, pixel-to-pixel variations
were removed using spectra of the internal lamp.  For the long
slit spectra, sky emission for each \ion{H}{ii} region was
subtracted by identifying a neighbouring area without emission.
For the multi-object spectra, two adjacent holes were always used,
one for the object (\ion{H}{ii} region or standard star) and the
other for the sky, which ensured the same spectral range for each
object-sky pair. The standard star spectra were obtained using the
masks cut for the \ion{H}{ii} regions.  The spectra of all objects
were calibrated in wavelength using the arc lamp spectra obtained
at the same time as the observations. Finally, the flux
calibration was made using the observations of the
spectrophotometric standard stars listed in Table \ref{Obslog} to
obtain the instrumental sensitivity function.

\subsection{Photometry}

\begin{figure}
\resizebox{\hsize}{!}{\includegraphics{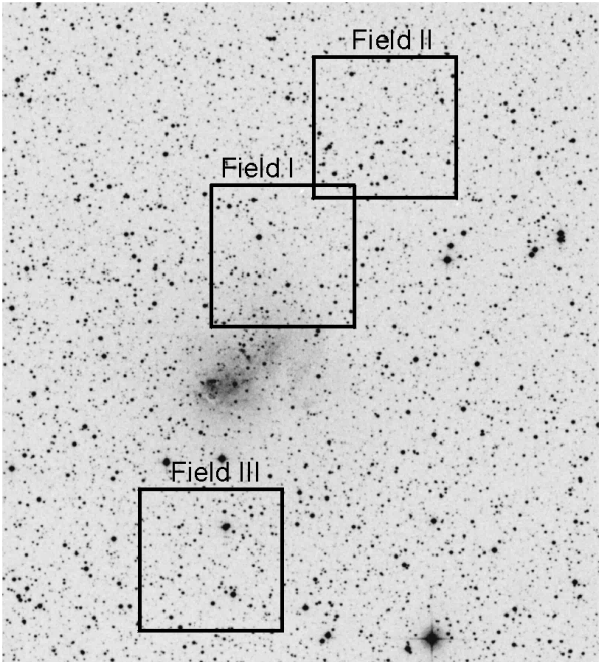}}
\vspace{-2.5cm}
\caption{This is an STScI Digitized Sky Survey
image of IC 10 ($B$-band) showing the locations of the fields for
which we obtained optical photometry.  North is up and east is to
the left.  The boxes indicate the $5.\!\arcmin 6
\times 5.\!\arcmin 6$ field of view of the CCD images.  The central
coordinates of each field are (epoch J2000): Field I: $0^{\mathrm
h}20^{\mathrm m}20^{\mathrm s}$, $59^{\circ}22^{\mathrm
m}25^{\mathrm s}$; Field II: $0^{\mathrm h}19^{\mathrm
m}30^{\mathrm s}$, $59^{\circ}26^{\mathrm m}43^{\mathrm s}$; and
Field III: $0^{\mathrm h}20^{\mathrm m}25^{\mathrm s}$,
$59^{\circ}10^{\mathrm m}39^{\mathrm s}$.}
\label{fields}
\end{figure}

The optical photometry of foreground fields towards IC 10 was
obtained at the Bulgarian National Astronomical Observatory
(Rozhen). The CCD had a plate scale of $0.\!\arcsec 33\,{\rm
pixel}^{-1}$, yielding a $5.\!\arcmin 6 \times 5.\!\arcmin 6$
field of view. The seeing was $\sim 1.\!\arcsec 2$ under stable
and very good photometric conditions. Three selected fields
towards IC 10 were observed, and are shown in Fig. \ref{fields}.
Sets of $UBVR$ images were obtained for Fields I and II, and a
$BVR$ set was obtained for Field III. Landolt (\cite{Landolt1992})
standards were taken before and after all observations.  The IRAF
data reduction package was used to carry out the basic image
reductions. The stellar photometry of the frames was performed
with the point-spread function fitting routine ALLSTAR available
in DAOPHOT (Stetson \cite{Stetson1993}).  Complete details of the
data reduction and analysis may be found in Georgiev et al.
(\cite{Georgievetal1999}).

\section{A new WR star}

\begin{figure}
\resizebox{\hsize}{!}{\includegraphics{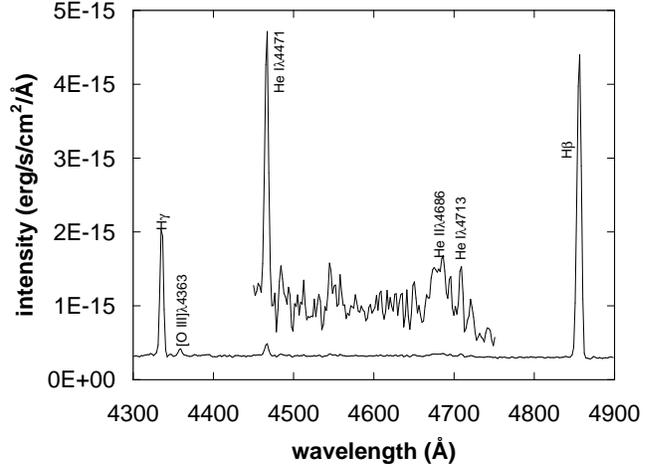}}
\caption{ The Steward spectrum of HL111c showing the broad
\ion{He}{ii}$\lambda$4686 bump.  The intensity scale is that for
the full spectrum.}
\label{wrbump}
\end{figure}

The Steward spectrum of HL111c (Fig. \ref{wrbump}) contained broad
\ion{He}{ii}$\lambda$4686 emission indicating the presence of at
least one WR star (WN specifically since there is no \ion{C}{iv}
emission at 4650\AA). This feature is absent in the SPM spectrum,
undoubtedly due to the slightly different placement of the slit
for the two observations.  In accordance with the stronger
continuum expected due to the WR star, the equivalent widths of
all of the emission lines are smaller in the Steward spectrum. The
\ion{He}{ii}$\lambda$4686 emission has an equivalent width of
3\AA.  Given the correlation between \ion{He}{ii}$\lambda$4686 and
H$\beta$ line intensities in WN stars, we expect an H$\beta$
equivalent width of order 0.6\AA\ in emission (Conti et al.
\cite{Contietal1983}).

\section{Line intensities}

\begin{table}
\caption[]{Raw line intensity ratios from Steward for the \ion{H}{ii}
regions in IC 10$\,^{\mathrm a}$} \label{Stewardint}
\[
\begin{tabular}{lccc}
\hline \noalign{\smallskip} $\lambda$& HL106b    & HL111b           & HL111c\\
\noalign{\smallskip}
\hline
\noalign{\smallskip}
{[}\ion{O}{ii}{]}\hfill 3727   &$  90.3\pm  7.9 $&$  142.9\pm  4.2 $&$  56.1\pm  1.5 $\\
H9\hfill 3835                  &$               $&$    2.9\pm  1.4 $&$  1.82\pm 0.75 $\\
{[}\ion{Ne}{iii}{]}\hfill 3869 &$               $&$    5.3\pm  1.8 $&$  12.3\pm  1.1 $\\
H8\hfill 3889                  &$               $&$    6.6\pm  1.9 $&$   8.4\pm  1.1 $\\
H$\epsilon$\hfill 3970         &$   9.2\pm  3.6 $&$    8.9\pm  2.4 $&$  11.5\pm  1.2 $\\
H$\delta$\hfill 4102           &$  14.8\pm  4.2 $&$   12.7\pm  1.8 $&$ 15.23\pm 0.96 $\\
H$\gamma$\hfill 4340           &$  29.6\pm  3.4 $&$   31.9\pm  1.5 $&$ 34.75\pm 0.54 $\\
{[}\ion{O}{iii}{]}\hfill 4363  &$   3.8\pm  2.0 $&$   1.91\pm 0.89 $&$  1.86\pm 0.34 $\\
\ion{He}{i}\hfill 4472         &$               $&$    3.8\pm  1.3 $&$  3.57\pm 0.40 $\\
H$\beta$\hfill 4861            &$ 100.0\pm  3.4 $&$  100.0\pm  1.5 $&$100.00\pm 0.74 $\\
{[}\ion{O}{iii}{]}\hfill 4959  &$ 124.1\pm  6.5 $&$   63.5\pm  2.6 $&$ 139.0\pm  3.3 $\\
{[}\ion{O}{iii}{]}\hfill 5007  &$ 386.8\pm  9.5 $&$  195.0\pm  3.5 $&$ 414.6\pm  4.5 $\\
$E(B-V)$                       &$  0.99\pm 0.28 $&$   0.77\pm 0.12 $&$  0.67\pm 0.03 $\\
$N(\ion{H}{i})/10^{20}$        & 15.1            & 13.1             & 2.31            \\
$T_e$ $(10^4\,\mathrm{K})$     & $1.4\pm 0.4$    & $1.4\pm 0.3$     & $1.00\pm 0.06$  \\
$12+\log(\mathrm O/\mathrm H)$ & $7.86\pm 0.32$  & $7.84\pm 0.25$   & $8.23\pm 0.09$  \\
\noalign{\smallskip}
\hline
\end{tabular}
\]
\begin{list}{}{}
\item[$^{\mathrm a}$] The object identifications are from Hodge \&
Lee (\cite{HodgeLee1990}).
\end{list}
\end{table}

Tables \ref{Stewardint}-\ref{Cananeaint} list the raw line
intensities and reddenings for the \ion{H}{ii} regions observed in
IC 10. The line intensities were measured using locally-developed
software (McCall et al. \cite{McCalletal1985}). The errors quoted
for each line intensity include contributions from the fit to the
line itself, from the fit to the reference line, and from the
noise in the continuum for both the line and reference line. Table
\ref{Stewardint} also includes the electron temperature computed
from the [\ion{O}{iii}] lines and the corresponding oxygen
abundance, calculated using the prescription given in Richer
(\cite{Richer1993}) but adopting an electron density of
$100\,\mathrm{cm}^{-3}$.

To compute the reddening, we first corrected the Balmer lines for
underlying stellar absorption according to
\begin{equation}
$$F_c(\lambda) = F_o(\lambda) \times
\left(1+\frac{W_\mathrm{abs}}{W_\lambda}\right) /
\left(1+\frac{W_\mathrm{abs}}{W_{\mathrm H \beta}}\right)
\label{eqnwidth}$$,\end{equation} where $F_o(\lambda)$ and
$F_c(\lambda)$ are the observed and corrected fluxes and
$W_\mathrm{abs}$, $W_\lambda$, and $W_{\mathrm H \beta}$ are the
equivalent widths of the underlying stellar absorption, of the
emission line in question, and of H$\beta$, respectively.  In all
cases, except for the Steward spectrum of HL111c, we used
$W_\mathrm{abs} = 1.9$\AA\ (McCall et al. \cite{McCalletal1985}).
Actually, our H$\alpha$-based reddenings would not be changed
significantly for any $W_\mathrm{abs} < 5$\AA. For the Steward
spectrum of HL111c, $W_\mathrm{abs}$ was required to be less than
0.2\AA\ in order to derive an H$\gamma$-based reddening that is
consistent with the H$\alpha$- and H$\gamma$-based reddenings from
the SPM spectrum, i.e., there is very little remaining absorption
once one accounts for the H$\beta$ emission from the WR star
(0.6\AA). The values given in Table \ref{Stewardint} were
calculated using $W_\mathrm{abs} = 0$\AA. Though an upper limit of
0.8\AA\ is considerably smaller than the typical Balmer absorption
found by McCall et al. (\cite{McCalletal1985}), it is not
unreasonable for a very young cluster (Santos et al.
\cite{Santosetal1995}).

We used these corrected line intensities to calculate the
reddening via
\begin{equation}
$$\log \frac{F(\lambda)}{F(\mathrm H\beta)} = \log
\frac{I(\lambda)}{I(\mathrm H\beta)} - 0.4
E(B-V)\left(A_1(\lambda) - A_1(\mathrm H\beta)\right)
\label{eqnredd}$$,
\end{equation}
where $F(\lambda)$ and $I(\lambda)$ are the observed and
theoretical emission-line fluxes at wavelength $\lambda$, $E(B-V)$
is the reddening, and $A_1(\lambda) = A(\lambda)/E(B-V)$.  Values
of $A_1(\lambda)$ were derived from Fitzpatrick's
(\cite{Fitzpatrick1999}) monochromatic reddening law parametrized
with a ratio of total-to-selective extinction of 3.041.  This
parametrization gives a law that delivers a true ratio of
total-to-selective extinction equal to 3.07 when integrated over
the spectrum of Vega (McCall \& Armour \cite{McCallArmour2000}).
For all of the H$\alpha$- and H$\gamma$-based reddenings of
\ion{H}{ii} regions in IC 10, we adopted $I({\mathrm
H}\alpha)/I({\mathrm H}\beta)= 2.86$ and $I({\mathrm
H}\gamma)/I({\mathrm H}\beta)= 0.468$, respectively, appropriate
for an electron density and temperature of $100\,\mathrm
{cm}^{-3}$ and $10^4\,{\mathrm K}$, respectively (Osterbrock
\cite{Osterbrock1989}).

Several external tests indicate that our flux calibrations, and
hence our reddenings, are secure within the quoted errors.  The
$\mathrm H\gamma$ and H$\delta$ line intensities from the Steward
and SPM long slit spectra (Tables \ref{Stewardint} and
\ref{SPMint}) are identical within errors for all objects in
common.  (Blueward of H$\delta$, the sensitivity of the SPM data
drops rapidly.)  All of the line intensities for HL111c from both
datasets agree with those observed by Lequeux et al.
(\cite{Lequeuxetal1979}). For the Cananea multi-object spectra
(Table \ref{Cananeaint}), the line intensities and reddening
we derive for HL45 are identical to those found by Lequeux et al.
(\cite{Lequeuxetal1979}).  There are no extant data with which we
might compare our Cananea long slit data (Table
\ref{Cananeaint}).  As we show below, however, the reddenings
we derive for HL77 and HL111d+e are consistent with those for the
other \ion{H}{ii} regions.  We therefore conclude that our flux
calibrations and the resulting reddenings are not affected by any
serious systematic errors.

\begin{table*}
\caption[]{Raw line intensity ratios from San Pedro M\'artir for the
\ion{H}{ii} regions in IC 10$\,^{\mathrm a}$} \label{SPMint}
\[
\begin{tabular}{lccccc}
\hline \noalign{\smallskip} $\lambda$  & HL106a & HL106b & HL111b & HL111c & HL111e \\
\noalign{\smallskip}
\hline
\noalign{\smallskip}
{[}\ion{O}{ii}{]}\hfill   3727 &$             $&$                $&$    184 \pm  64 $&$     86 \pm  16 $&$     85 \pm  31 $\\
{[}\ion{Ne}{iii}{]}\ \    3869 &$             $&$                $&$                $&$    22.8\pm 8.0 $&$                $\\
H$\delta$\hfill           4102 &$             $&$     28 \pm  12 $&$    14.1\pm 6.6 $&$    13.2\pm 3.3 $&$                $\\
H$\gamma$\hfill           4340 &$  29 \pm  10 $&$     29 \pm  11 $&$    29.8\pm 7.2 $&$    33.6\pm 2.0 $&$    33.2\pm 5.2 $\\
H$\beta$\hfill            4861 &$ 100 \pm  12 $&$    100 \pm  11 $&$   100.0\pm 7.2 $&$   100.0\pm 2.2 $&$   100.0\pm 4.7 $\\
{[}\ion{O}{iii}{]}\hfill  4959 &$ 61.6\pm 8.4 $&$   116.5\pm 9.4 $&$    72.5\pm 6.1 $&$   127.2\pm 4.4 $&$   106.8\pm 6.5 $\\
{[}\ion{O}{iii}{]}\hfill  5007 &$ 194 \pm  10 $&$    362 \pm  11 $&$   218.5\pm 7.0 $&$   384.2\pm 5.1 $&$   308.0\pm 7.5 $\\
\ion{He}{i}\hfill         5876 &$ 24.1\pm 9.7 $&$    28.2\pm 7.1 $&$    21.4\pm 4.3 $&$    22.5\pm 1.7 $&$    20.4\pm 3.3 $\\
{[}\ion{O}{i}{]}\hfill    6300 &$             $&$                $&$                $&$    5.2 \pm 1.3 $&$    10.2\pm 2.4 $\\
{[}\ion{N}{ii}{]}\hfill   6548 &$  27 \pm  11 $&$    13.8\pm 9.2 $&$    21.6\pm 6.2 $&$    9.0 \pm 5.0 $&$    12.4\pm 4.2 $\\
H$\alpha$\hfill           6563 &$ 911 \pm  18 $&$    1093\pm  14 $&$   767.2\pm 8.2 $&$   627.4\pm 8.8 $&$   695.7\pm 5.9 $\\
{[}\ion{N}{ii}{]}\hfill   6583 &$ 109 \pm  14 $&$     76 \pm  11 $&$    84.0\pm 6.6 $&$    30.9\pm 6.2 $&$    56.2\pm 4.7 $\\
\ion{He}{i}\hfill         6678 &$ 12.9\pm 6.5 $&$    5.7 \pm 6.9 $&$    5.7 \pm 2.9 $&$    6.9 \pm 1.1 $&$    5.7 \pm 2.5 $\\
{[}\ion{S}{ii}{]}\hfill   6716 &$ 128 \pm  11 $&$    54.6\pm 9.5 $&$    72.3\pm 4.9 $&$    23.9\pm 1.3 $&$    53.8\pm 3.9 $\\
{[}\ion{S}{ii}{]}\hfill   6731 &$  91 \pm  10 $&$    47.1\pm 9.3 $&$    53.6\pm 4.6 $&$    19.0\pm 1.3 $&$    36.6\pm 3.7 $\\
\ion{He}{i}\hfill         7065 &$ 14.3\pm 6.3 $&$    20.2\pm 6.8 $&$                $&$    5.3 \pm 1.2 $&$    7.9 \pm 2.6 $\\
{[}\ion{Ar}{iii}{]}\hfill 7136 &$ 27.3\pm 7.8 $&$    52.5\pm 8.5 $&$                $&$    28.4\pm 1.5 $&$    31.0\pm 3.5 $\\
{[}\ion{O}{ii}{]}\hfill   7325 &$             $&$    19.2\pm 8.5 $&$    16.0\pm 4.9 $&$                $&$    63.4\pm 4.3 $\\
$E(B-V)$                       &$1.02 \pm 0.12$&$    1.17\pm 0.10$&$    0.85\pm 0.07$&$    0.68\pm 0.02$&$    0.77\pm 0.04$\\
$N(\ion{H}{i})/10^{20}$        & 15.5          & 15.1             & 13.1             & 2.31             & 3.08             \\
\noalign{\smallskip}
\hline
\end{tabular}
\]
\begin{list}{}{}
\item[$^{\mathrm a}$] The object identifications are from Hodge \&
Lee (\cite{HodgeLee1990}).
\end{list}
\end{table*}

\begin{table*}
\caption[]{Raw line intensity ratios from Cananea for the
\ion{H}{ii} regions in IC 10$\,^{\mathrm a}$}
\label{Cananeaint}
\[
\begin{tabular}{lcccccc}
\hline \noalign{\smallskip}
technique & multi-object & multi-object & multi-object &
multi-object & long slit & long slit \\
$\lambda$ & HL30 & HL45 & HL50 & HL81 & HL111d+e & HL77 \\
\noalign{\smallskip} \hline
H$\delta$\hfill          4102 &$                  $&$    9.1\pm  2.0 $&$   16.7\pm  5.3 $&$                $\\
H$\gamma$\hfill          4340 &$       49\pm   17 $&$   29.6\pm 1.4  $&$   34.5\pm  3.8 $&$                $&$   24.9\pm  8.3 $&$     32\pm   13 $\\
{[}\ion{O}{iii}{]}\hfill 4363 &                    &                  &                  &                  &$                $&$   15.6\pm  9.3 $\\
\ion{He}{i}\hfill        4472 &$                  $&$    3.0\pm 0.95 $&$    5.9\pm  2.5 $&$                $&                  &                  \\
H$\beta$\hfill           4861 &$      100\pm   16 $&$  100.0\pm 2.4  $&$  100.0\pm  4.0 $&$    100\pm   28 $&$    100\pm   11 $&$    100\pm   15 $\\
{[}\ion{O}{iii}{]}\hfill 4959 &$       56\pm   13 $&$  187.5\pm  4.0 $&$  112.6\pm  5.5 $&$    136\pm   28 $&$    125\pm   13 $&$    143\pm   16 $\\
{[}\ion{O}{iii}{]}\ \    5007 &$      162\pm   19 $&$  562.2\pm  8.6 $&$    374\pm   10 $&$    416\pm   61 $&$    365\pm   20 $&$    401\pm   27 $\\
\ion{He}{i}\hfill        5876 &$     31.5\pm  5.5 $&$  28.02\pm 0.73 $&$   17.7\pm  1.9 $&$                $&$   23.9\pm  9.9 $&$                $\\
{[}\ion{N}{ii}{]}\hfill  6548 &$                  $&$                $&$    3.1\pm   10 $&$     45\pm   23 $\\
H$\alpha$\hfill          6563 &$      947\pm   63 $&$    891\pm   13 $&$    682\pm   19 $&$    887\pm  122 $&$    747\pm   31 $&$    702\pm   41 $\\
{[}\ion{N}{ii}{]}\hfill  6583 &$       81\pm   13 $&$   29.8\pm  3.7 $&$   16.5\pm  6.8 $&$     54\pm   19 $&$   42.1\pm  6.8 $&$   45.3\pm  9.6 $\\
\ion{He}{i}\hfill        6678 &$                  $&$  11.12\pm 0.63 $&$    8.4\pm  1.3 $&$                $\\
{[}\ion{S}{ii}{]}\hfill  6716 &$       81\pm   12 $&$  31.51\pm 0.87 $&$   23.1\pm  1.8 $&$    103\pm   23 $&$   31.0\pm  7.3 $&$   34.9\pm  9.7 $\\
{[}\ion{S}{ii}{]}\hfill  6731 &$       56\pm   11 $&$  28.43\pm 0.83 $&$   16.8\pm  1.7 $&$     41\pm   17 $&$   27.2\pm  7.1 $&$   31.1\pm  9.4 $\\
\ion{He}{i}\hfill        7065 &$                  $&$  11.56\pm 0.56 $&$    6.1\pm  1.1 $&$                $\\
{[}\ion{Ar}{iii}{]}\hfill 7136&$                  $&$                $&$   27.8\pm  1.5 $&$                $\\
$E(B-V)$                      &$   1.06 \pm 0.07  $&$ 1.00 \pm 0.01  $&$  0.77 \pm 0.03 $&$  0.94 \pm 0.15 $&$  0.84 \pm 0.04 $&$  0.79 \pm 0.05 $\\
$N(\ion{H}{i})/10^{20}$       & 7.48               & 2.43             & 7.30             & 5.91             & 3.08             & 4.10             \\
\hline
\end{tabular}
\]
\begin{list}{}{}
\item[$^{\mathrm a}$] The object identifications are
from Hodge \& Lee (\cite{HodgeLee1990}).
\end{list}
\end{table*}

\section{The foreground reddening towards IC 10}

\begin{figure}
\resizebox{\hsize}{!}{\includegraphics{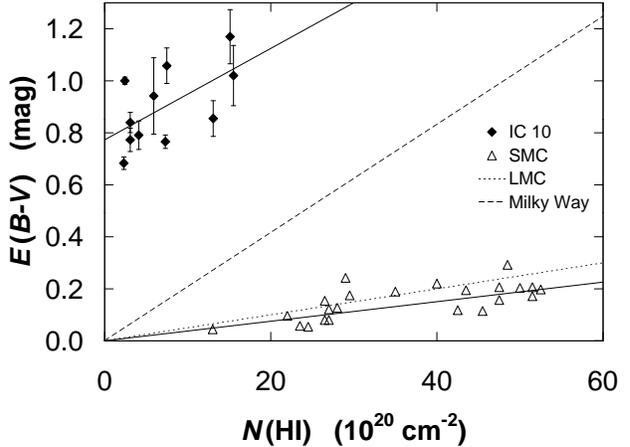}}
\caption{ The
reddening as a function of the \ion{H}{i}
column density.  For IC 10 and the SMC,
the reddening towards each \ion{H}{ii} region is plotted as a
function of half of the total column
density along the line of sight within either IC 10 or the SMC.
Were there no
reddening due to the Milky Way, the fit to the data for IC 10 and
the SMC (solid lines) should pass through the
origin. For IC 10, the vertical intercept yields a foreground
reddening due to dust within the Milky Way of $E(B-V)=0.77 \pm
0.07\,\mathrm{mag}$. For the SMC, the fit implies a foreground
reddening of $0.02\pm 0.03\,\mathrm{mag}$, in agreement
with other studies (e.g., Schlegel et al. \cite{Schlegeletal1998}).
The error bars on the IC 10 points are those given in Tables
\ref{Stewardint}-\ref{Cananeaint}.
Also shown are the relationships between $E(B-V)$ and
$N(\ion{H}{i})$ for the LMC and the Milky Way (Koornneef
\cite{Koorneef1982}; Bohlin et al. \cite{Bohlinetal1978}).}
\label{ebvnh1}
\end{figure}

The minimum reddening towards an \ion{H}{ii} region in IC 10 is a
first estimate of the foreground reddening towards this galaxy.
The minimum value in Tables \ref{Stewardint}-\ref{Cananeaint} is
$0.68\pm 0.02\,\mathrm{mag}$ for the \ion{H}{ii} region HL111c.
(Throughout we use H$\alpha$-based reddenings when available.)
This value is substantially lower than the values found in the
literature, commonly $0.85-1.15\,{\mathrm {mag}}$ (cf. Sect. 6).
The reddening towards HL111c is all the more surprising since this
\ion{H}{ii} region likely suffers some reddening due to dust
within IC 10.

A second estimate of the foreground reddening toward IC 10 may be
obtained from a plot of the reddening for each \ion{H}{ii} region
as a function of the column density of \ion{H}{i} within IC 10
along that line of sight (McCall \cite{McCall1989}).  The
\ion{H}{i} in the central part of IC 10 is in a disk (Shostak \&
Skillman \cite{ShostakSkillman1989}; Wilcots \& Miller
\cite{WilcotsMiller1998}) and the \ion{H}{ii} regions we observed
are necessarily embedded within it.  Thus, in Fig. \ref{ebvnh1},
we plot the reddening as a function of half of the \ion{H}{i}
column density along the line of sight within IC 10 using the
Wilcots \& Miller (\cite{WilcotsMiller1998}) column density map
(11\arcsec\ beam; their Fig. 2). The adopted column densities are
given in Tables \ref{Stewardint}-\ref{Cananeaint} and were
determined from a surface brightness map provided by B. Miller. As
can be seen, there is a good correlation. The straight line is an
unweighted least squares fit to all of the data points. This fit
extrapolates to a reddening of $0.77\pm 0.07\,{\mathrm {mag}}$ at
zero column density. We adopt this value as our second estimate
for the foreground reddening toward IC 10.

As a test of this method, in Fig. \ref{ebvnh1} we also plot the
reddenings and \ion{H}{i} column densities for the \ion{H}{ii}
regions in the SMC from Caplan et al. (\cite{Caplanetal1996}). We
excluded their data for N84c and N88, which deviate most strongly.
The fit to the remaining points yields a foreground reddening of
$0.02\pm 0.03\,\mathrm{mag}$.  The Schlegel et al.
(\cite{Schlegeletal1998}) maps indicate reddenings of
$0.03-0.05\,\mathrm{mag}$ in the vicinity of the SMC. Likewise,
McNamara \& Feltz (\cite{McNamaraFeltz1980}) found a foreground
reddening of $0.02\,\mathrm{mag}$.  That our intercept is in good
agreement with these values implies that the \ion{H}{i} is
reliably tracking the dust content. Thus, the intercept appears to
yield the true foreground reddening.

The scatter about the fitted line for IC 10 in Fig. \ref{ebvnh1},
$0.13\,\mathrm{mag}$, is likely primarily due to variations in
foreground dust within the Milky Way. This scatter could be
produced by fluctuations of order 10\% of the mean foreground
\ion{H}{i} column density, $6.4 \times
10^{21}\,\mathrm{atoms}/\mathrm{cm}^2$ (Cohen \cite{Cohen1979}),
supposing the gas-to-dust ratio found by Bohlin et al.
(\cite{Bohlinetal1978}).  If due entirely to dust within IC 10,
the scatter would require differences in the depths of the
\ion{H}{ii} regions within IC 10's disk of order $\pm 26\%$ of the
total column density (Shostak \& Skillman
\cite{ShostakSkillman1989}) or might be due to unresolved dust
clouds. Regardless of the origin of the scatter, its magnitude is
easily accounted for and its effect upon the intercept in Fig.
\ref{ebvnh1} should be minimal by virtue of our large number of
sight lines.

Given IC 10's low galactic latitude, $b = -3^\circ$, it was not
removed from the Schlegel et al. (\cite{Schlegeletal1998})
reddening maps. Indeed, at the position of IC 10, there is a
feature of roughly the size expected for IC 10.  Although the area
around IC 10 in these maps is very complex, the foreground
reddening implied is $E(B-V) = 0.75-1.00\,\mathrm{mag}$, in
excellent agreement with our spectroscopy of IC 10's \ion{H}{ii}
regions. Within IC 10, the reddening maps indicate
$E(B-V)_\mathrm{max} = 1.60\,\mathrm{mag}$.

de Vaucouleurs \& Ables (\cite{deVaucouleursAbles1965}) deduce a
foreground reddening of $E(B-V)=0.87\pm 0.05\,\mathrm{mag}$ by
comparing IC 10's apparent colours with those for unreddened dwarf
irregulars.  If IC 10's colours are atypical of dwarf irregulars
(Sect. 7), this method will over-estimate the reddening.

\begin{figure}
\resizebox{\hsize}{!}{\includegraphics{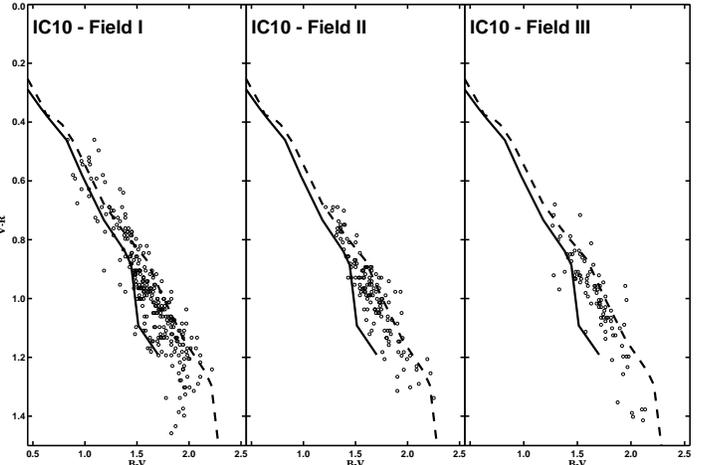}}
\smallskip
\caption{The $(B-V)$ vs. $(V-R)$ diagrams for Fields I, II, and
III. Superposed are colour-colour lines for dwarfs (Bessell
\cite{Bessell1990}). The solid line is the colour-colour relation
in the absence of reddening, whereas the dashed line corresponds
to our adopted reddening towards IC 10 of
$E(B-V)=0.77\,\mathrm{mag}$. Clearly, these foreground stars
provide only a lower limit to the reddening towards IC 10.  For
all of the stars shown here, the formal DAOPHOT errors are smaller
than $0.15\,\mathrm{mag}$. } \label{u-bb-v}
\end{figure}

Finally, in Fig. \ref{u-bb-v}, we present colour-colour diagrams
based upon our $BVR$ photometry for the foreground stars along the
line of sight towards IC 10 in Fields I, II, and III (Fig.
\ref{fields}).  We fit the colours of main sequence stars (Bessell
\cite{Bessell1990}) to our $(B-V)-(V-R)$ colour-colour diagrams to
obtain a mean reddening for the stars in all three fields of
$E(B-V)=0.37\pm 0.06\,\mathrm{mag}$, with no evidence for
significant variation between them.  These foreground stars will
not sample the full column density of gas and dust towards IC 10,
so this estimate of the foreground reddening should be interpreted
as a lower limit to the reddening towards IC 10.

Considering the above evidence as well as that to be presented in
the following section, we adopt the intercept in Fig.
\ref{ebvnh1}, $E(B-V)=0.77\pm 0.07\,\mathrm{mag}$, as the most
reliable estimate of the foreground reddening toward IC 10 since
it is based upon many independent reddening estimates along
individual lines of sight.

\section{Internal dust}

Fig. \ref{ebvnh1} clearly implies that IC 10 contains a
considerable amount of internal dust.  The slope found differs
from zero at the 95\% confidence limit, despite the scatter.
Extant studies of IC 10's different stellar populations
corroborate the existence of internal dust. Massey \& Armandroff
(\cite{MasseyArmandroff1995}) obtain $E(B-V)=0.80\pm
0.05\,\mathrm{mag}$ and $0.75\,\mathrm{mag}$ for WR and blue plume
stars, respectively. Sakai et al. (\cite{Sakaietal1999}) obtain
$E(B-V)=1.16\pm 0.08\,\mathrm{mag}$ and $0.85\,\mathrm{mag}$ for
Cepheids and RGB stars, respectively. Borissova et al.
(\cite{Borissovaetal2000}) find $E(B-V)=1.05\pm
0.10\,\mathrm{mag}$ and $1.8\pm 0.2\,\mathrm{mag}$ for red
supergiant and Br$\gamma$ sources, respectively.  Generally,
younger stellar populations suffer greater reddening. The
exceptions are the blue plume and WR reddenings, but these are
likely skewed towards the lower limits for these stellar
populations since the least reddened stars in these populations
are the easiest to study. Curiously, the reddening found for
Br$\gamma$ sources is similar to the maximum reddening in the
Schlegel et al. (\cite{Schlegeletal1998}) maps. All of these
reddenings match or exceed our foreground value, implying the
reality of significant internal dust.

The slope of the relation in Fig. \ref{ebvnh1} is
$E(B-V)/N(\ion{H}{i}) = (1.7\pm 0.8) \times
10^{-22}\,\mathrm{mag}\,\mathrm{cm}^2$, only slightly less than
that found for the Milky Way (Knapp \& Kerr \cite{KnappKerr1974};
Bohlin et al. \cite{Bohlinetal1978}; Burstein \& Heiles
\cite{BursteinHeiles1978}).

\section{The nature of IC 10}

\begin{table}
\caption[]{The properties of IC 10} \label{photprop}
\[
\begin{tabular}{lcc}
\hline \noalign{\smallskip} Observed & & Source \\ \hline
\noalign{\smallskip} $B_T$ & $11.80 \pm 0.20\,\mathrm{mag}$ & 1 \\
$(B-V)$ & $1.30\pm 0.03\,\mathrm{mag}$ & 2 \\
$(U-B)$ & $0.25\pm 0.03\,\mathrm{mag}$ & 2 \\
$\mu_e(B)$ & $24.09\,\mathrm{mag}/\sq\arcsec$ & 2 \\
$r_e$ & $2.\!\arcmin 0\pm 0.\!\arcmin 2$ & 2 \\
$(m - M)$ & $24.1 \pm 0.2\,\mathrm{mag}$ & 3 \\
$E(B-V)_{\mathrm{gal}}$ & $0.77\pm 0.07\,\mathrm{mag}$ & 4 \\
\noalign{\smallskip} \hline \noalign{\smallskip} Adopted & & \\
\noalign{\smallskip} \hline\noalign{\smallskip} $R_V$ & 3.05 & 5 \\
\noalign{\smallskip} \hline \noalign{\smallskip} Derived & & \\
\noalign{\smallskip} \hline\noalign{\smallskip}
$B_T^0$   & $8.68\pm 0.42\,\mathrm{mag}$ & \\
$(B-V)^0$ & $0.53\pm 0.10\,\mathrm{mag}$ & \\
$(U-B)^0$ & $-0.32\pm 0.08\,\mathrm{mag}$ & \\
$\mu_e^0(B)$ & $20.97\pm 0.36\,\mathrm{mag}/\sq\arcsec$ & \\
$\mu_e^0(V)$ & $20.44\pm 0.38\,\mathrm{mag}/\sq\arcsec$ & \\
$r_e$ & $0.38\pm 0.05\,\mathrm{kpc}$ & \\
$M_B$ & $-15.42\pm 0.46\,\mathrm{mag}$ & \\
\noalign{\smallskip} \hline
\end{tabular}
\]
References: (1) RC3: de Vaucouleurs et al. \cite{rc3}; (2) de
Vaucouleurs \& Ables \cite{deVaucouleursAbles1965}; (3) Sakai et
al. \cite{Sakaietal1999}; (4) this work; (5) McCall \& Armour
\cite{McCallArmour2000};
\end{table}

In light of IC 10's foreground reddening, we now consider its
photometric properties, and, in particular, whether it should be
considered a typical dwarf irregular. We summarize some of IC 10's
relevant properties in Table \ref{photprop}.

There is no modern surface photometry for IC 10.  The best
available photoelectric aperture photometry is that of de
Vaucouleurs \& Ables (\cite{deVaucouleursAbles1965}), who obtained
the results listed in Table \ref{photprop}.  Since their data for
their smallest apertures is uncertain, we concentrate on their
effective aperture values, i.e., the values for an aperture
containing one half of IC 10's total luminosity.

We adopt the distance modulus from Sakai et al.
(\cite{Sakaietal1999}), which is based upon $VIJHK$ photometry of
Cepheids, incorporates a self-consistent reddening, and is
corroborated by that found by Borissova et al.
(\cite{Borissovaetal2000}).

The derived photometric parameters were calculated as follows:

$B_T^0 = B_T - A_B$,

$(B-V)^0 = (B-V) - E(B-V)$,

$(U-B)^0 = (U-B) - E(U-B)$,

$\mu_e^0(B) = \mu_e(B) -A_B$,

$\mu_e^0(V) = \mu_e^0(B) - (B-V)^0$

\noindent where

$A_B = (1+R_V)E(B-V)$,

\noindent and

$E(U-B)/E(B-V) = 0.70 + 0.05\,E(B-V)$

\noindent (FitzGerald \cite{FitzGerald1970}).

The colours we derive for IC 10 are rather different from those of
dwarf irregulars (de Vaucouleurs et al.
\cite{deVaucouleursetal1981}).  Although its $(U-B)$ colour is
similar to that of dwarf irregulars, its $(B-V)$ colour is much
too red. Instead, IC 10's colours are very similar to those of
Markarian galaxies (Huchra \cite{Huchra1977}).

\begin{figure}
\resizebox{\hsize}{!}{\includegraphics{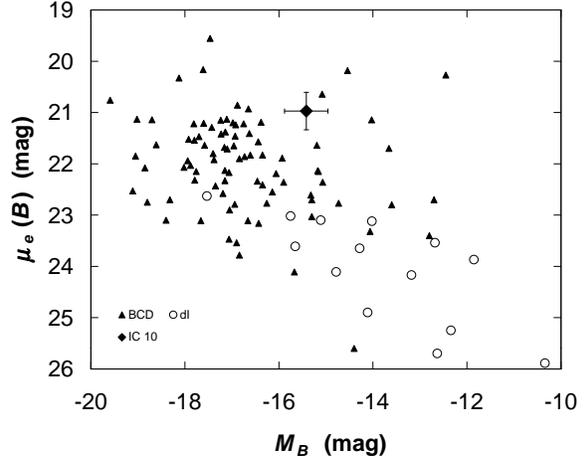}}
\caption{ The
$\mu_e(B)$--$M_B$ diagram for star-forming dwarf galaxies.  The
dwarf irregular data is from the Bremnes et al.
(\cite{Bremnesetal1998}, \cite{Bremnesetal1999}) photometry of the
M81 and M101 groups, using distance moduli of
$27.80\,\mathrm{mag}$ and $29.30\,\mathrm{mag}$ respectively
(Freedman et al. \cite{Freedmanetal1994}, Jurcevic et al.
\cite{Jurcevicetal2000}).  The BCD data are from Papaderos et al.
(\cite{Papaderosetal1996}) and the Vennik et al.
(\cite{Venniketal2000}) local samples. }
\label{mbmueff}
\end{figure}

The surface brightnesses we derive for IC 10 are very high!
Admittedly, the higher reddening values mentioned earlier imply
even higher surface brightnesses, but we believe our foreground
reddening is more reliable and the photometric parameters we
derive therefore more reliable. Patterson \& Thuan
(\cite{PattersonThuan1996}) find that the {\it central} surface
brightness for typical dwarf irregulars is fainter than
$22.7\,\mathrm{mag}/\sq\arcsec$ in the $B$ band, nearly two
magnitudes fainter than the mean surface brightness within IC 10's
effective radius.  In fact, IC 10's effective surface brightness
is brighter than the canonical central surface brightness for
spiral galaxy disks (Freeman \cite{Freeman1970}).

Only when we consider more actively star-forming galaxies do we
find surface brightnesses similar to that of IC 10.  Fig.
\ref{mbmueff} illustrates this graphically. Here, we plot the mean
effective surface brightness as a function of absolute magnitude,
both in the $B$ band, for various samples of dwarf galaxies.  IC
10's effective surface brightness is typical of that of BCDs, but
at least two magnitudes brighter than that of typical dwarf
irregulars.  Quantitatively, IC 10's effective surface brightness
is similar to or brighter than the mean for the galaxy samples
plotted in Fig. \ref{mbmueff} or of the mean for the dwarfs in
Telles \& Terlevich's (\cite{TellesTerlevich1997}) sample of
\ion{H}{ii} galaxies. IC 10's effective radius is typical of
galaxies in any of these samples.

Massey \& Armandroff (\cite{MasseyArmandroff1995}) found at least
15 WR stars in IC 10.  We have located another in our Steward
spectrum of HL111c.  Thus, at least 16 WR stars are known within
the $7\arcmin \times 5\arcmin\!.5$ extent of the optically-bright
part of the galaxy.  Given IC 10's distance of $661\,\mathrm{kpc}$
(Sakai et al. \cite{Sakaietal1999}), its surface density of WR
stars is then $11.2\,\mathrm{kpc}^{-2}$. This surface density
exceeds that found in discrete centres of active star formation in
local galaxies by at least a factor of two (Massey \& Johnson
\cite{MasseyJohnson1998}), even though it applies to the entire
area of the optically-bright part of the galaxy. Thus, there can
be little doubt of the exceptional strength of IC 10's recent
episode of star formation.

IC 10's $12\mu\mathrm{m}/25\mu\mathrm{m}$ flux ratio is consistent
with its dust emission being dominated by a strong starburst
component (IRAS PSC \cite{PSC}; Helou \cite{Helou1986}). The
$60\mu\mathrm{m}/100\mu\mathrm{m}$ flux ratio is less conclusive
because of the uncertainty in the $100\mu\mathrm{m}$ flux.

In summary, IC 10 is not a typical dwarf irregular.  Given the
foreground reddening we obtain, which is lower than most of the
values that have been suggested, its surface brightness is typical
of that found for dwarf starburst galaxies.  The same is true of
the scale length of the optically-bright part of the galaxy.  That
the entire surface area of the optically-bright part of the galaxy
has a surface density of WR stars double that in active centres of
massive star formation observed elsewhere leaves no doubt that IC
10 is undergoing an intense starburst on a large scale.  Thus, IC
10 should be considered to be a blue compact dwarf galaxy (BCD).


\begin{acknowledgements}
      We thank Margarita Vald\'ez Guti\'errez for her help in
      obtaining the telescope time at the Observatorio
      Astrof\'\i sico Guillermo Haro.  We thank Bryan Miller for
      allowing us the use of his \ion{H}{i} map in constructing
      Fig. \ref{ebvnh1}.  We thank the anonymous referee for
      helpful comments.
      MGR thanks Stanley Kurtz and Robin
      Fingerhut for very helpful discussions.
      MGR thanks the staff of the Observatorio Astrof\'\i sico
      Guillermo Haro for their hospitality and Ra\'ul Gonz\'alez and
      Gerardo Miramon for their able assistance with the
      observations.
      AB acknowledges a Mutis graduate fellowship from the
      Agencia Espa\~nola de Cooperaci\'on Internacional.
      MR, MGR, and AB
      acknowledge financial support from CONACYT
      project 27984-E and DGAPA project IN122298.
      Support for JB and RK was provided by a Bulgarian  National
      Science Foundation grant under contract No. F-812/1998 with
      the Bulgarian Ministry of Education and Sciences.
      MLM and RLK thank the staff of the Observatorio
      Astron\'omico Nacional in
      San Pedro M\'artir for its help with the observations
      and gratefully acknowledge the continuing support of
      the Natural Sciences and Engineering Research Council of
      Canada.  MLM thanks the staff of the Steward
      Observatory for its assistance with observations.
      LG acknowledges financial support from DGAPA project
      IN113999.
\end{acknowledgements}

\end{document}